\documentclass[prd,a4paper,aps,twocolumn,nofootinbib,nobibnotes,preprintnumbers]{revtex4} 

\usepackage[dvips]{graphicx}
\usepackage{amsmath,amssymb,mathrsfs}
\usepackage{bm}
\usepackage{times}
\usepackage{epsfig}
\usepackage{verbatim}
\usepackage{bm}
\usepackage[utf8]{inputenc}
\usepackage{graphics}
\usepackage{graphicx,epsfig,amssymb,amsmath,color, cancel}
\usepackage{soul}
\usepackage[normalem]{ulem}
\usepackage{slashed}
\usepackage{comment}

\usepackage[english]{babel}
\usepackage{amsfonts,amsmath,amssymb,amsthm,mathtools}
\usepackage{comment,enumerate,footnote,graphicx,subfloat,relsize}
\usepackage{array,tabularx,tabu,multirow,framed,afterpage}
\usepackage[usenames,dvipsnames]{xcolor}
\usepackage{cleveref}
\crefformat{pluralequation}{#2\black{eqs.~(}#1\black{)}#3}
\Crefformat{pluralequation}{#2\black{Equations~(}#1\black{)}#3}
\crefformat{pluralfigure}{#2\black{figs.~}#1#3}
\Crefformat{pluralfigure}{#2\black{Figures~}#1#3}

\newcommand{\be}{\begin{equation}}
\newcommand{\ee}{\end{equation}}

\begin{document}

\title{
Leptogenesis with perturbations in type-II and type-III seesaw models 
}

\author{Iason Baldes}
\email{iasonbaldes@gmail.com}
\affiliation{Laboratoire de Physique de l'\'Ecole Normale Sup\'erieure, ENS, \\ Universit\'e PSL, CNRS, Sorbonne Universit\'e, Universit\'e Paris Cit\'e, F-75005 Paris, France}


\begin{abstract}
Density perturbations have recently been shown to lead to a novel effect in the freeze-out of heavy particles called ``acoustically driven freeze-out." This leads to an enhancement in the yield in standard leptogenesis. We extend this calculation to include $2 \to 2$ washout processes in type-I leptogenesis and the Sommerfeld-enhanced $2 \to 2$ gauge annihilations in type-II and type-III leptogenesis. These CP conserving gauge annihilations suppress the yield of heavy particles sourcing the asymmetry. We show the acoustically driven freeze-out leads to novel enhancements in the baryon asymmetry in type-II and type-III leptogenesis, already in the weak washout regime, in contrast with type-I leptogenesis.
\end{abstract}

\maketitle

\section{Introduction}

Cosmological inflation, or an early universe phase transition, may have led to enhanced density perturbations at small scales. When these perturbations enter the Hubble horizon they undergo oscillations which also leads to oscillations in the temperature. Freeze-out processes play a key role in early universe cosmology, both in well known standard model processes, such as the free-electron fraction during recombination and the neutron abundance during big bang nucleosynthesis, and also in speculative beyond the standard model processes, such as dark matter freeze out and baryogenesis through leptogenesis~\cite{Dodelson:2003ft,Peter:2013avv}. 

Recently, Hotokezaka, Jinno, and Takada, have shown an interesting non-linear effect arises when taking into account the possibility of temperature oscillations during freeze-out processes, which they dubbed ``acoustically driven freeze-out"~\cite{Hotokezaka:2025ewq}. The effect arises due to the Boltzmann factor in the number density of an in-equilibrium heavy particle, of mass $M$, which given a perturbation in the temperature, $T = \bar{T} + \delta T$, behaves as
	\begin{align}
	\mathrm{exp}\left( -M/T \right) & = \mathrm{exp}\left( -M/ (\bar{T} + \delta T) \right) \nonumber \\
					& = \mathrm{exp}\left( -M/\bar{T} \right)\mathrm{exp}\left( M \delta T/\bar{T}^2 \right). \label{eq:bmannsup}
	\end{align}
The key observation being that the $\mathrm{exp}\left(M \delta T/\bar{T}^2 \right)$ factor may well be large in the Boltzmann suppressed regime, as the $M/\bar{T} \gg 1$ enhancement can overcome the $\delta T/\bar{T} \lesssim 1$ suppression. This leads to important corrections to freeze-out processes. Studying type-I leptogenesis as an example, Hotokezaka, Jinno, and Takada showed the baryon yield to be enhanced by $\sim (10 - 30)\%$, in the strong washout regime, for Horizon sized density perturbations with primordial curvature perturbation $\mathcal{R}_{i} \sim (0.1 - 0.3)$, after averaging over the phase of the oscillation. The reason the effect is most relevant in the strong washout regime, is that that is where the decay-and-inverse-decay interplay is most important in keeping the abundance of heavy neutrinos close to the equilibrium value during the time the asymmetry is being generated. Thus the argument using the formula in Eq.~\eqref{eq:bmannsup} is properly applicable only in the strong washout regime, which was furthermore shown using a detailed numerical calculation. (For earlier work focusing on small amplitude perturbations, see~\cite{Kartavtsev:2008fp}.)

To be of interest for acoustically driven freeze-out, the length scale of the density perturbation should also be of Horizon size, $\sim 1/H(\bar{T})$, a little before freeze-out, i.e.~at $\bar{T} \sim M$. The reason is that super-horizon size perturbations will average out over each Hubble patch which is effectively acting a separate universe at that epoch. Furthermore, perturbations over length scales much smaller than the Horizon size will be oscillating very rapidly, so the effect during freeze-out will also be averaged out. That leaves only the sweet-spot of perturbations of length $\sim 1/H$ to play a significant role. 

In type-I leptogenesis~\cite{Fukugita:1986hr}, the abundance of the heavy Majorana neutrinos is controlled by their decay rate, and therefore one power of the Boltzmann factor enters in the calculation. In many freeze-out processes, however, there are two heavy particles involved on the same side of the reaction, and therefore two powers of the Boltzmann factor will enter. Rather obviously, we then have
		\begin{align}
	\mathrm{exp}\left( -2M/T \right) & = \mathrm{exp}\left( -2M/ (\bar{T} + \delta T) \right) \nonumber \\
					& = \mathrm{exp}\left( -2M/\bar{T} \right)\mathrm{exp}\left( 2M \delta T/\bar{T}^2 \right),  \label{eq:bmannsup2}
	\end{align}
and therefore one can speculate that the correction will tend to be larger (due to the well established mathematical inequality $2 > 1$). This is relevant for standard DM self-annihilations during freeze-out, and also for type-II~\cite{Ma:1998dx,Hambye:2003ka,Antusch:2004xy,Hambye:2005tk} and type-III leptogenesis~\cite{Hambye:2003rt,Strumia:2008cf,Hambye:2012fh}. In this paper, we study the latter two leptogenesis models as examples of acoustically driven freeze-out involving annihilations, and leave DM freeze-out for another day~\cite{baldes2} (annihilations can also play an important role in simple extensions of the type-I scenario~\cite{AristizabalSierra:2014uzi}, but we only consider the minimal scenario here).

As we shall see below, in our detailed analysis, the size of the effect for the annihilations is of the same order as for the decays, once the full dynamics is taken into account. (Rather than larger as the factor of two in Eq.~\eqref{eq:bmannsup2} above may indicate.) Nevertheless, the effect comes into play in different areas of parameter space of type-II and type-III leptogenesis, compared to the type-I example, and therefore offers a novelty.  

One objection to this whole enterprise may be that without a reason for expecting significant enhancements in $\mathcal{R}_{i}$ --- at length scales of order of the hubble horizon $\sim 1/H$ during the freeze-out process --- these calculations are only of academic interest, and will be unlikely to be of relevance for explaining the observed universe, unlike the perturbations at CMB scales~\cite{Senatore:2008vi} (for historical studies of small scale inhomogeneties during BBN, see e.g.~\cite{Terasawa:1988aa,Kurki-Suonio:1989vaa}, and for more recent studies of the effect of density perturbations on BBN see~\cite{Jeong:2014gna,Nakama:2014vla,Inomata:2016uip}). We note, however, that strong phase transitions associated with the generation of heavy DM~\cite{Hambye:2013dgv,Hambye:2018qjv,Baldes:2018emh,Baldes:2020kam,Baldes:2021aph,Kierkla:2022odc,Wong:2023qon,Balan:2025uke} or heavy (seesaw) states associated with baryogensis~\cite{Lazarides:1985ja,Buchmuller:2012wn,Buchmuller:2013lra,Brivio:2017dfq,Brivio:2018rzm,Brdar:2018vjq,Brdar:2018num,Brdar:2019iem,Brivio:2019hrj,Brdar:2019qut,Baldes:2021vyz,Huang:2022vkf,Dasgupta:2022isg,Chun:2023ezg}, will also lead to the generation of curvature perturbations with length scales $\sim 1/H$~\cite{Liu:2022lvz,Elor:2023xbz,Lewicki:2024ghw,Zou:2025sow,Franciolini:2025ztf}. If the reheating temperature is subsequently high enough for the DM or heavy seesaw states to come back into equilibrium, the generated perturbations may well then lead to acoustically driven freeze-out playing a role.

This motivates the current study in which we further explore acoustically driven freeze-out in isolation from other effects. The calculation in a more complete scenario can then benefit from the understanding gained. Similarly, the role of all these phase transitions, inhomogeneities, and density perturbations in generating detectable gravitational waves as a further signal of such a scenario is left for future work.

Finally, we note that the isocurvature perturbations on observable CMB scales --- which are super-horizon sized when sourced from out-of-equilibrium effects in our leptogenesis scenario, or from dark-matter freeze-out/in --- will be highly suppressed and so not constrain the scenario~\cite{Weinberg:2004kf,Bellomo:2022qbx,Strumia:2022qvj,Racco:2022svs,Holst:2023msh,Stebbins:2023wak}.

\section{Leptogenesis models}

Perhaps the simplest way to explain the neutrino masses inferred from oscillation data~\cite{Davis:1968cp,SNO:2001kpb,Super-Kamiokande:1998kpq,K2K:2006yov,MINOS:2011qho,DayaBay:2012fng,RENO:2012mkc} are the tree-level seesaw models. The standard type-I seesaw features heavy right handed sterile neutrinos~\cite{Minkowski:1977sc,Yanagida:1979as,Gell-Mann:1979vob,Mohapatra:1979ia}, transforming trivially $N \sim (1,1,0)$ under the SM gauge group, the type-II seesaw uses a scalar triplet $T \sim (1,3,2)$~\cite{Schechter:1980gr,Lazarides:1980nt,Mohapatra:1980yp,Wetterich:1981bx}, and the type-III seesaw fermion triplets $\Sigma \sim (1,3,0)$~\cite{Foot:1988aq}. Upon integrating out the heavy states, all lead to the standard Weinberg operator $(LH)^2/\Lambda$ at low energies, giving the neutrino masses.

One has the following interactions for type I
	\begin{equation}
	\mathcal{L} \supset -y_{N} \tilde{H}^{\dagger} \bar{N} L - y_{N}^{\dagger} \tilde{H} \bar{L} N ,
	\end{equation}
which lead to the decays $N \to HL$ and $N \to \bar{H}\bar{L}$ into SM Higgs and lepton doublets. For type-II, along with the gauge interactions, we also have
	\begin{equation}
	\mathcal{L} \supset  -\frac{1}{2}(y_{\mathsmaller{TL}} L^{T} C i \tau_{2} T L + y_{\mathsmaller{TH}} M_{T} \tilde{H}^{T} i \tau_{2} T \tilde{H}+\mathrm{H.c.}),
	\end{equation}
where $M_{T}$ is the triplet mass, which give the decays $T \to \bar{L}\bar{L}$ and $T \to HH$. Finally, for type-III, we again have gauge interactions and in addition
	\begin{equation}
	\mathcal{L} \supset  -\sqrt{2} y_{\Sigma} \tilde{H}^{\dagger} \bar{\Sigma} L - \sqrt{2} y_{\Sigma}^{\dagger} \tilde{H} \bar{L} \Sigma ,
	\end{equation}
which lead to the decays $\Sigma \to HL$ and $\Sigma \to \bar{H}\bar{L}$. In all cases, for precise definitions and technical details, see~\cite{Hambye:2012fh}. 

In all three of the above seesaw realizations, a $\mathcal{B}-\mathcal{L}$ asymmetry can be generated through CP violating decays of the heavy states, in a rather straight forward fashion. We consider the decays of the lightest of the heavy states, with mass $M_{1}$, and parameterize the total decay rate in the standard way. We write the effective neutrino mass as~\cite{Buchmuller:2004nz,Hambye:2005tk}
	\begin{align}
	\tilde{m_{1}} & = \frac{(y_{N}y_{N}^{\dagger})_{11} v^{2} }{ M_{1} }, \quad \mathrm{[Type-I]}, \\
	\tilde{m_{1}} & = \sqrt{(y_{\mathsmaller{TH}} y_{\mathsmaller{TH}}^{\dagger})  \mathrm{Tr} (y_{\mathsmaller{TL}} y_{\mathsmaller{TL}}^{\dagger}) } \frac{ v^{2} }{ M_{1} }, \quad \mathrm{[Type-II]}, \\ 
	\tilde{m_{1}} & = \frac{(y_{\Sigma}y_{\Sigma}^{\dagger})_{11} v^{2} }{ M_{1} }, \quad \mathrm{[Type-III]},
	\end{align}
where $v = 174$ GeV is the electroweak VEV. For the type-I and type-III seesaws we have the decay rate
	\begin{align}
	\Gamma_{N} & = \frac{ \tilde{m_{1}}M_{1}^2 }{ 8\pi v^{2} }, \quad \mathrm{[Type-I]}, \\
	\Gamma_{\Sigma} & = \frac{ \tilde{m_{1}}M_{1}^2 }{ 8\pi v^{2} }, \quad \mathrm{[Type-III]}.
	\end{align}
For the type-II scenario we have two decay channels, with decay rates~\cite{Hambye:2005tk}
	\begin{align}
	\Gamma(T \to \bar{L}\bar{L}) & \equiv B_{L}\Gamma_{T}  = \mathrm{Tr} (y_{\mathsmaller{TL}} y_{\mathsmaller{TL}}^{\dagger}) \frac{ M_{1} }{16 \pi} \\
	\Gamma(T \to HH) & \equiv B_{H}\Gamma_{T}  = (y_{\mathsmaller{TH}} y_{\mathsmaller{TH}}^{\dagger}) \frac{ M_{1} }{16 \pi},
	\end{align}
where $B_{L}+B_{H}=1$ are the branching fractions. One thus has
	\begin{equation} 
	\Gamma_{T} = \frac{ \tilde{m_{1}}M_{1}^2 }{ 16\pi v^{2} \sqrt{B_L B_H  } }, \quad \mathrm{[Type-II}].
	\end{equation}
For simplicity, when we scan over parameter space, solving the Boltzmann equations below, we will be considering the limit $B_L=B_H=0.5$. Then
	\begin{equation}
	\Gamma_{T} = \frac{ \tilde{m_{1}}M_{1}^2 }{ 8\pi v^{2} }, \quad \mathrm{[Type-II, \; Simplified]}.
	\end{equation}
Generically, we then write $\Gamma_{D} \equiv \Gamma_{N} = \Gamma_{\Sigma} = \Gamma_{T} $, and the standard decay parameter is given by
	\begin{equation}
	K \equiv \frac{ \Gamma_{D} }{ H(T=M_1) } =  \frac{ \tilde{m_{1}} }{ m_{\ast} },
	\end{equation}
where $m_{\ast} \simeq 1.1 \times 10^{-3} \; \mathrm{eV}$ is the equilibrium neutrino mass~\cite{Buchmuller:2004nz}. This distinguishes the weak, $K<1$, and strong washout regimes, $K > 1$.

The CP violation is parameterized in a standard way:
	\begin{align}
	\epsilon_{N} & \equiv \frac{ \Gamma(N \to HL) - \Gamma(N \to \bar{H}\bar{L}) } {\Gamma(N \to HL) + \Gamma(N \to \bar{H}\bar{L}) }, \quad \mathrm{[Type-I]}, \\
	\epsilon_{T} &\equiv 2 \frac{ \Gamma(\bar{T} \to LL) - \Gamma(T \to \bar{L}\bar{L}) }{ \Gamma_{\bar{T}} + \Gamma_{T} }, \quad \mathrm{[Type-II]}, \\
	\epsilon_{\Sigma} & \equiv \frac{ \Gamma(\Sigma \to HL) - \Gamma(\Sigma \to \bar{H}\bar{L}) } {\Gamma(\Sigma \to HL) + \Gamma(\Sigma \to \bar{H}\bar{L}) }, \quad \mathrm{[Type-III]}.
	\end{align}
In the limit of hierarchical spectrum of heavy states, there is a well known bound on the CP violation depending on $M_{1}$~\cite{Hambye:2001eu,Davidson:2002qv}, known as the Davidson-Ibarra bound in the type-I scenario, and which generalizes to type-II and type-III~\cite{Hambye:2003rt,Hambye:2005tk}. One has~\cite{Davidson:2002qv,Hambye:2005tk,Hambye:2003rt}
	\begin{align}
	|\epsilon_{N}| & < \frac{3}{16\pi} \frac{M_{1}}{v^{2}} (m_{\nu 3} - m_{\nu 1}), \quad \mathrm{[Type-I]}, \label{eq:cptype1} \\
	|\epsilon_{T}| & < \frac{1}{4\pi} \frac{ M_{1} }{ v^{2} } \sqrt{ B_{L} B_{H}  \sum m_{\nu i}^{2} }, \quad \mathrm{[Type-II]}, \label{eq:cptype2} \\
	|\epsilon_{\Sigma }| & < \frac{1}{16\pi} \frac{M_{1}}{v^{2}}(m_{\nu 3} - m_{\nu 1}), \quad \mathrm{[Type-III]}, \label{eq:cptype3}
	\end{align}
where $m_{\nu 3} - m_{\nu 1}$ is the difference between the heaviest and lightest of the active neutrinos. We take $m_{\nu 3} - m_{\nu 1} \approx 0.1 \; \mathrm{eV}$ and $\sum m_{\nu i}^{2} = (0.1 \; \mathrm{eV})^{2}$ throughout this paper.
We will use these bounds to demarcate the most natural parameter regions for successful leptogenesis in the usual manner. 

\section{Density and temperature perturbations}

The perturbations we consider are identical to those used by Hotokezaka, Jinno, and Takada~\cite{Hotokezaka:2025ewq} and consist of radiation density perturbations oscillating in a radiation dominated universe. We follow their definitions and take $T = \bar{T} + \delta T$, where $\bar{T}$ is the spatially averaged temperature of the universe, and define the fluctuation as $\delta_T \equiv \delta T/\bar{T}$. As a ``time" variable  in the Boltzmann equations it is useful to define  
	\begin{equation}
	\bar{z} \equiv \frac{ M_{1} }{ \bar{T} }.
	\end{equation}
Furthermore, it is also useful to introduce
	\begin{equation}
	z_{T} \equiv  \frac{ M_{1} }{ T } =  \frac{ \bar{z} }{ 1 + \delta_T }.
	\end{equation}
We work in Newtonian gauge of linearized cosmological perturbation theory, with negligible anisotropic stress, so that the two scalar perturbations of the metric are equal $\Phi = \Psi$. In radiation domination, for a given Fourier mode with wavenumber $k = | \mathbf{k}|$, one has~\cite{Peter:2013avv,Hotokezaka:2025ewq}
	\begin{equation}
	\Psi(\bar{z}) = 2 | \mathcal{R}_{i} | \cos{\delta} \, \frac{ \sin \varphi - \varphi \cos \varphi }{ \varphi^3 },
	\label{eq:Psifunc}
	\end{equation}  
where $\delta$ is a phase, and
	\begin{equation}
	\varphi(\bar{z}) = \frac{ k \eta }{ \sqrt{3} } \equiv \frac{ \bar{z} }{ \bar{z}_H },
	\end{equation}
where $\eta$ is the conformal time and $\bar{z}_H$ specifies the time of Horizon entry for the given wavenumber. We also require the temperature fluctuation
	\begin{equation}
	\delta_{T}(\bar{z}) = 2 | \mathcal{R}_{i} | \cos{\delta} \, \frac{ \sin \varphi - \varphi \cos \varphi - \varphi^{2} \sin \varphi + \frac{1}{2} \varphi^{3} \cos \varphi }{ \varphi^{3} }.
	\label{eq:deltafunc}
	\end{equation}
We are now ready to turn to the Boltzmann equations. 

\section{Asymmetry in the presence of perturbations}

\begin{figure*}[t]
\begin{center}
\includegraphics[width=0.32 \textwidth]{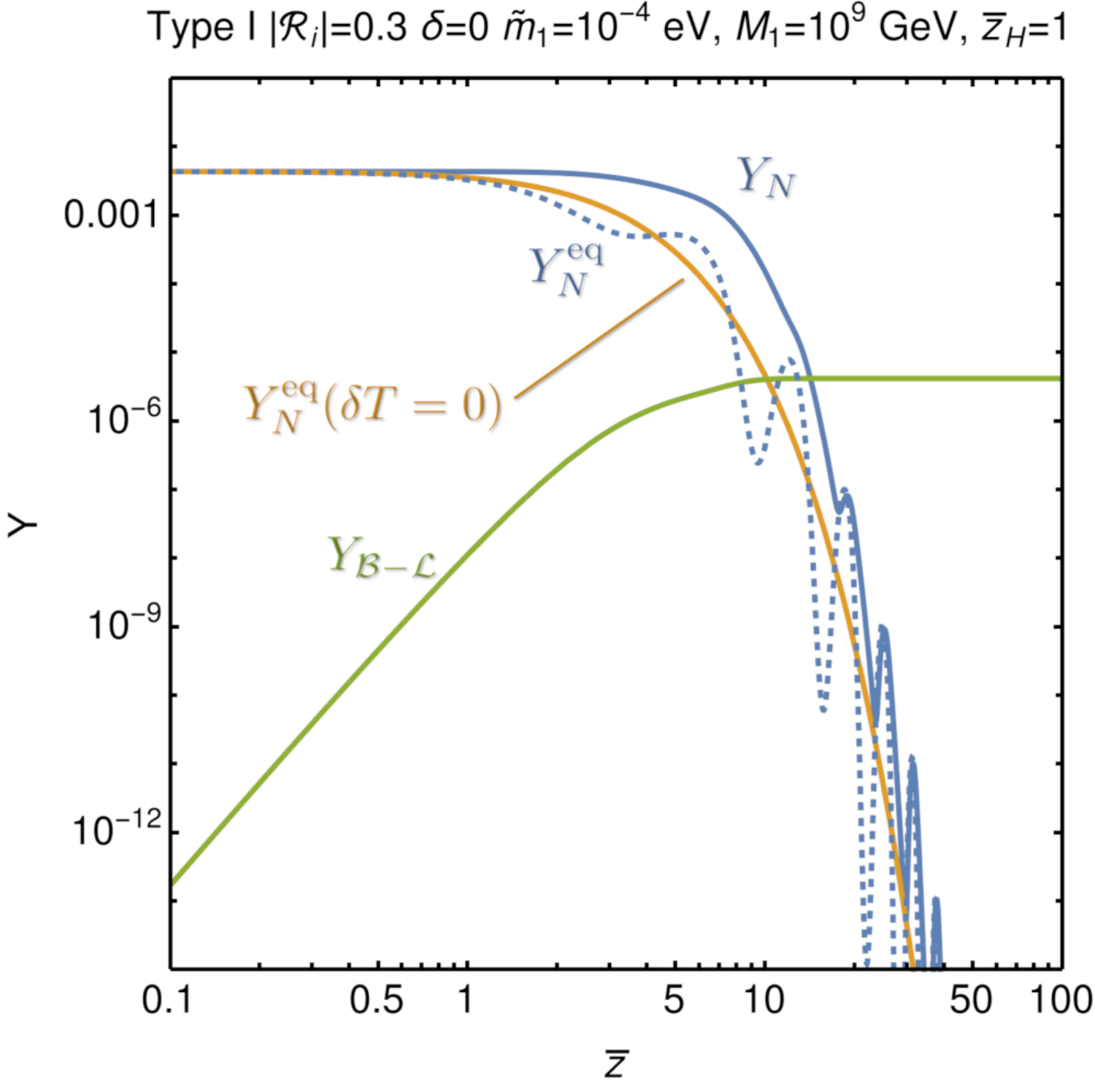} \,
\includegraphics[width=0.32 \textwidth]{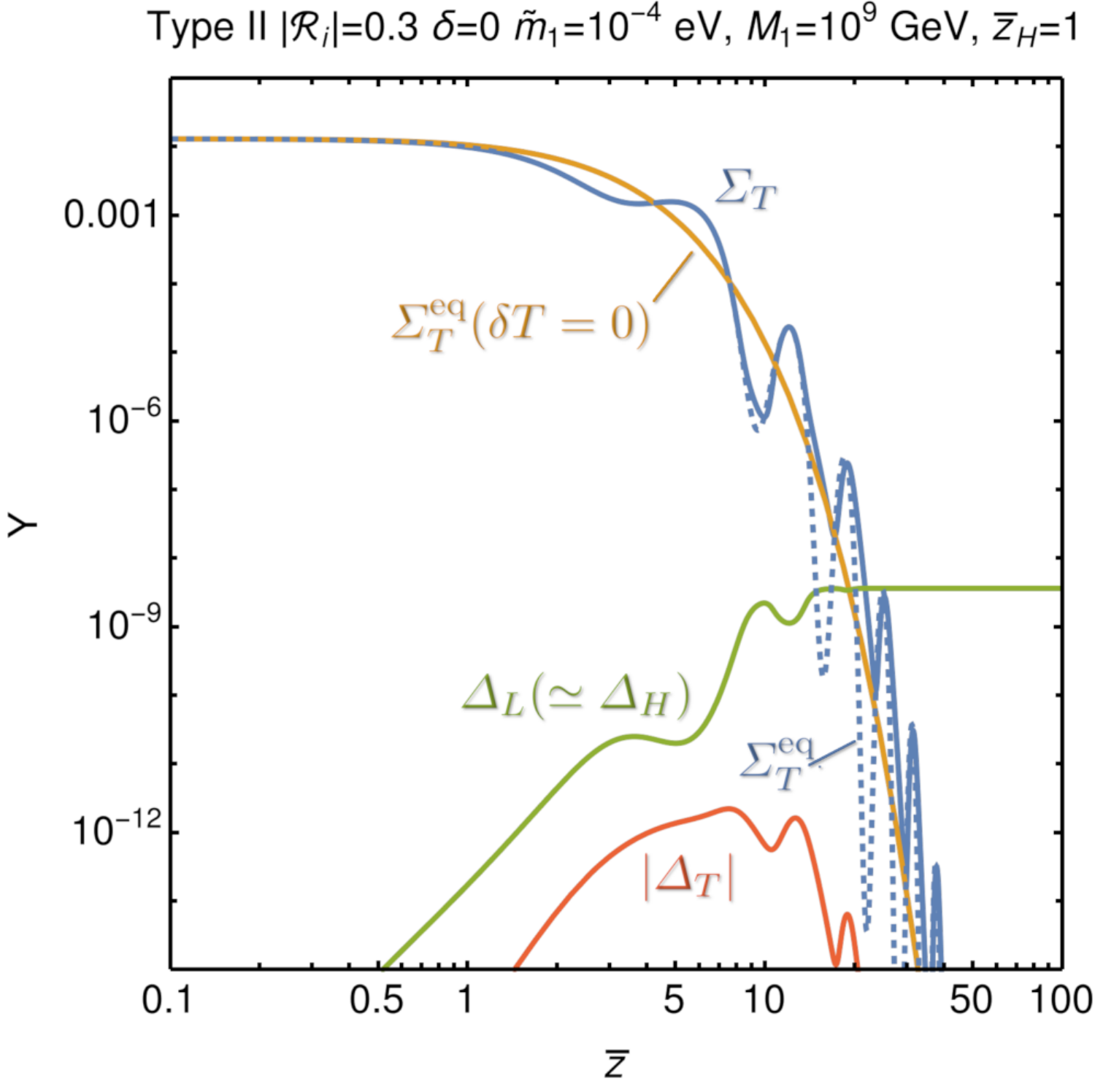} \,
\includegraphics[width=0.32 \textwidth]{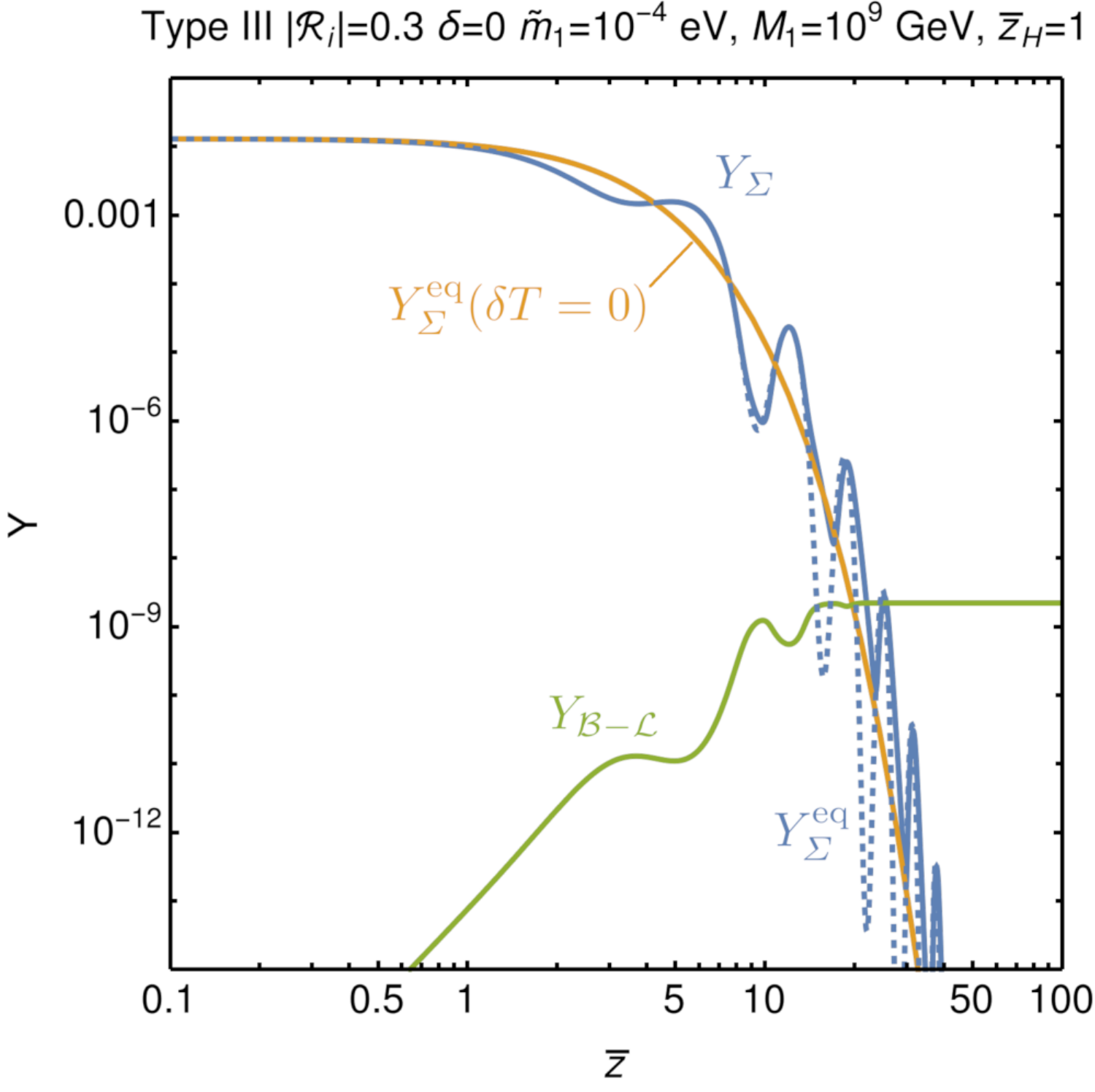}
\end{center}
\caption{
Example solutions to the Boltzmann equations in the weak washout regime for the three seesaws with $\epsilon_{N}=\epsilon_{T}=\epsilon_{\Sigma}=-10^{-3}$ (and $B_L=B_H=0.5$ for type-II). In the type-II and type-III seesaw the gauge interactions keep the number density of the decaying particle close --- but somewhat above --- the oscillating equilibrium value during the period of asymmetry generation. Thus the density perturbation leads to an enhanced yield for type-II and III in this example over their respective solutions when the perturbation is absent. Some other choices of $\delta$ lead to a suppression, but the overall average gives an enhancement.}
\label{fig:bmannexamples}
\end{figure*}	

\subsection{Type-I Boltzmann equation} 
We now give the Boltzmann equations for the seesaw models, following Hotokezaka, Jinno, and Takada~\cite{Hotokezaka:2025ewq}, in taking into account the perturbations. For type-I our results will necessarily match theirs, up to our inclusion of $2 \to 2$ washout processes, and a slight re-writing to aid comparison with previous literature relevant also for type-II and III. We begin with some useful definitions. We are interested in finding the Boltzmann equation for the evolution of the heavy Majorana neutrino density
	\begin{equation}
	Y_{N} = \frac{ n_{N} }{ s },
	\end{equation}
normalized to the entropy density,
 	\begin{equation}
 	 s(z_{T}) =  g_{\ast} \frac{ 2  \pi^{2} }{ 45 } \frac{M_{1}^{3} }{ z_{T}^{3} } ,
 	\end{equation}
\\
with $g_{\ast} = 106.75$ effective degrees-of-freedom. The baryon-minus-lepton asymmetry normalized to entropy is denoted $Y_{\mathcal{B} - \mathcal{L}}$.  The equilibrium number density of heavy neutrinos in the Maxwell-Boltzmann approximation is given by
	\begin{equation}
	Y_{N}^{\rm eq}(z_T) = \frac{ n_{N}^{\rm eq} }{ s } = \frac{ g_{N} M_{1}^3  }{ 2 \pi^2 z_T s(z_T)} K_{2}(z_{T})
	\end{equation}
where $g_{N} =2 $. The thermally averaged decay rate times number density is given by
	\begin{equation}
	\gamma_{N}(z_{T}) = \frac{ K_{1}(z_{T}) }{ K_{2}(z_{T}) } n_{N}^{\rm eq}(z_{T}) \Gamma_{N} . 
	\end{equation}
The washout rate depends on the distribution of the $\mathcal{B}-\mathcal{L}$ asymmetry into various particle species present in the plasma. For simplicity, we ignore detailed determination of such spectator processes~\cite{Buchmuller:2001sr,Nardi:2005hs} and also ignore flavour effects~\cite{Abada:2006fw,Nardi:2006fx,Abada:2006ea,Lavignac:2015gpa} in the present analysis, and consider simply an equilibrium number density of SM leptons, for which we take
	\begin{equation}
	Y_{L}^{\rm eq} =  \frac{ 135 \zeta(3) }{ 4 g_{\ast} \pi^4 },
	\end{equation}
corresponding simply to two massless fermionic degrees-of-freedom (we are counting the antiparticles separately). Note we are assuming the temperature oscillations are slow compared to equilibration times inside the SM plasma. 

Finally we also need the $2 \to 2$ washout rates. In the type-I seesaw we are interested in $LH \to \bar{H}\bar{L}$ and $LL \to \bar{H}\bar{H}$ scattering via $s-$channel and $t-$channel $N$ exchange. As inverse decays followed by decays of the on-shell $N$ are taken into account separately, we need to use the real intermediate state (RIS) subtracted s-channel propagator. For the purposes of the present paper, we use the formulas for the reduced cross sections for these processes, given as $\hat{\sigma}_{Ns}^{\rm sub}$ and $\hat{\sigma}_{Nt}$  in~\cite[Eqs.~(92) and (93)]{Giudice:2003jh}, massless SM particles, and the parameter $\xi = \mathrm{Max}[1, \sqrt{\Delta(m_{\rm atm})^{2}}/\tilde{m_{1}} ]$, where $\Delta(m_{\rm atm})^{2} \simeq 0.05 \; \mathrm{eV}$ which takes into account scatterings mediated by heavier right-handed neutrinos. The overall $2\to 2$ washout rate we then denote as
	\begin{equation}
	\gamma_{N}^{\rm sub}(z_T) =  \frac{1}{64\pi^{4} } \int ds s^{1/2} ( \hat{\sigma}_{Ns}^{\rm sub} + \hat{\sigma}_{Nt} ) K_{1}\left( \frac{ \sqrt{s} z_{T} }{ M_{1} } \right).
	\end{equation}
Having collected all these definitions and prior results, we can now report the Boltzmann equations. Aided by the fact that the collision term does not contain metric fluctuations in the so called Local Inertial Frame Instantaneously at Rest with respect to the Comoving Observer~\cite{Senatore:2008vi}, and following the careful derivation in~\cite{Hotokezaka:2025ewq}, we have:
\begin{widetext}
\begin{align}
[1 - \Psi(\bar{z})] s(z_{T})H(\bar{z})\bar{z} \frac{ d Y_{N} } { d \bar{z} } & = -  \gamma_{N}(z_{T}) \left[ \frac{Y_{N}}{Y_{N}^{\rm eq}(z_T) } - 1 \right], \\
[1 - \Psi(\bar{z})] s(z_{T})H(\bar{z})\bar{z} \frac{ d Y_{\mathcal{B}-\mathcal{L}} } { d \bar{z} } & = - \epsilon_{N} \gamma_{N}(z_{T}) \left[ \frac{Y_{N}}{Y_{N}^{\rm eq}(z_T) } - 1 \right] - \frac{ Y_{\mathcal{B}-\mathcal{L}} }{ Y_{L}^{\rm eq} } \left[ \frac{ \gamma_{N}(z_T) }{2} + 2 \gamma_{N}^{\rm sub}(z_T)   \right].
\end{align}
\\
\end{widetext}
One can readily check that, up to superficial choices of using entropy or photon density normalization, and the inclusion of $ \gamma_{N}^{\rm sub}$, these are equivalent to the equations reported in~\cite{Hotokezaka:2025ewq}. Compared to the usual Boltzmann equations without perturbations, the differences lie in the appearance of the perturbed temperature in the rates and equilibrium number densities, as well as the time dilation factor, $(1 - \Psi)$, on the LHS of the equations. After sphaleron reprocessing, the final baryon asymmetry is given by $  Y_{\mathcal{B}} = (28/79) \, Y_{\mathcal{B}-\mathcal{L}}$. We are ignoring corrections to the decay rate at NLO and scattering removing one $N$ from the thermal bath, which must be considered together to remove IR divergences, as these give only $\sim 5\%$ corrections to the $N$ decay rate~\cite{Salvio:2011sf}.

\begin{figure*}[t]
\begin{center}
\includegraphics[width=0.42 \textwidth]{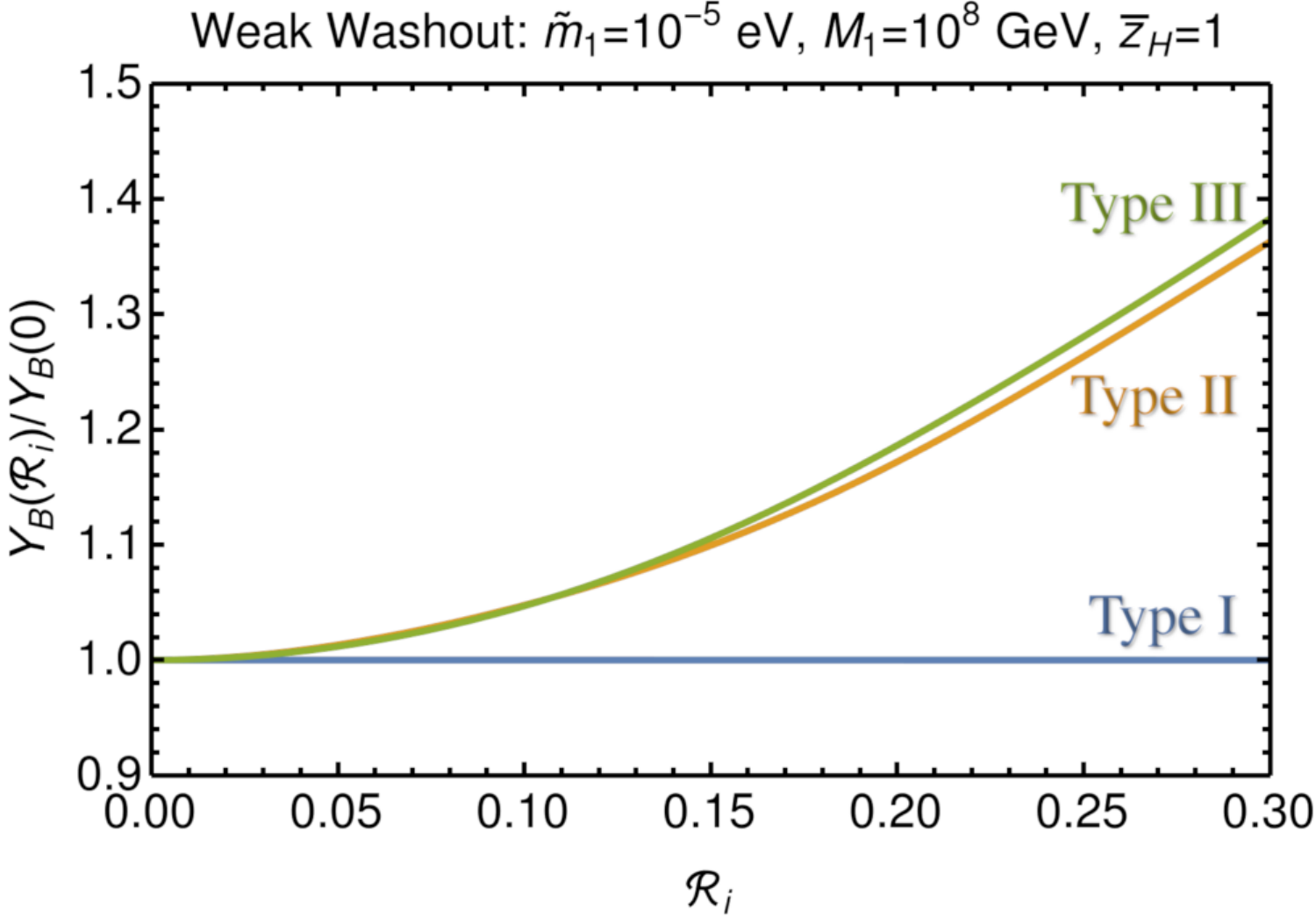} \qquad \qquad
\includegraphics[width=0.42 \textwidth]{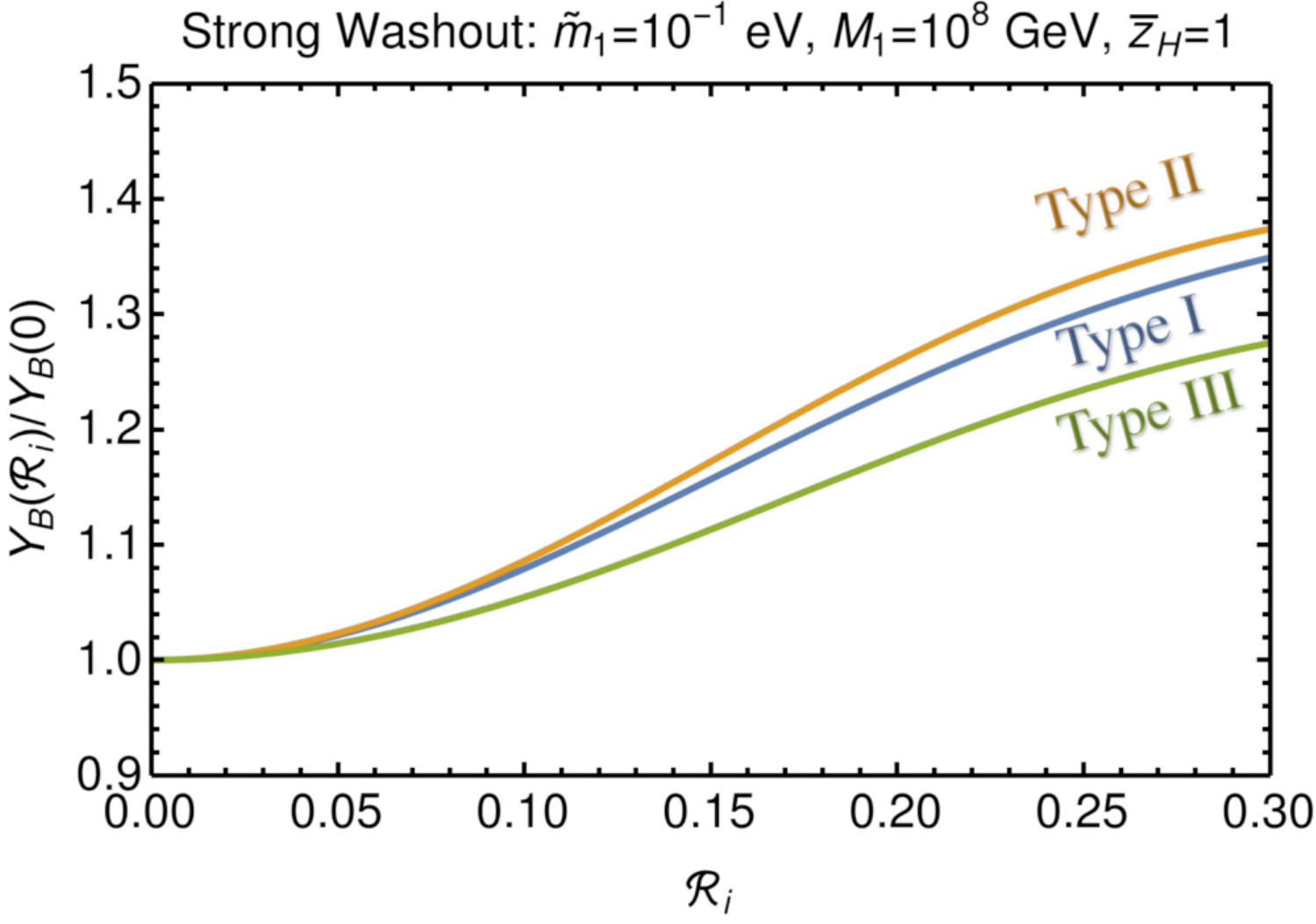}
\end{center}
\caption{Contrast of the enhancement achievable in the weak and strong washout regimes for the different leptogenesis scenarios where we have averaged over the phase $\delta$ ($B_L=B_H=0.5$ for type-II). The acoustically driven freeze-out leads to an enhancement already in the weak washout regime for type-II and III leptogenesis. Note there is also non-trivial dependence on $\tilde{m}_{1}$ and $M_{1}$ in all cases (see below). }
\label{fig:weakvsstrong}
\end{figure*}	
	
\subsection{Type-II Boltzmann equation} 

For type-II leptogenesis we begin by defining
	\begin{equation}
	\Sigma_{T} = \frac{n_{T}+ n_{\bar{T}}}{s},
	\end{equation}
with equilibrium number density
	\begin{equation}
	\Sigma_{T}^{\rm eq}(z_T) = \frac{n_{T}^{\rm eq}+ n_{\bar{T}}^{\rm eq}}{s} = \frac{ (g_{T}+g_{\bar{T}}) M_{1}^3  }{ 2 \pi^2 z_T s(z_T)} K_{2}(z_{T}),
	\end{equation}
where $g_{T}= g_{\bar{T}}=3$. The thermally averaged decay rate times number density is given by
	\begin{equation}
	\gamma_{T}(z_{T}) = \frac{ K_{1}(z_{T}) }{ K_{2}(z_{T}) } [n_{T}^{\rm eq}(z_{T})+ n_{\bar{T}}^{\rm eq}(z_{T})] \Gamma_{T} . 
	\end{equation}
We also require the asymmetries in the individual species, the BSM scalar triplet, the SM fermion doublet, and the SM Higgs doublet  
	\begin{equation}
	\Delta_{T} = \frac{ n_{T}-n_{\bar{T}} }{ s }, \;  \Delta_{L} = \frac{ n_{L}-n_{\bar{L}} }{ s } , \;  \Delta_{H} = \frac{ n_{H}-n_{\bar{H}} }{ s }. 
	\end{equation}
Note hypercharge conservation enforces 
	\begin{equation}
	2\Delta_T + \Delta_H - \Delta_L = 0.
	\end{equation}
We will also need the equilibrium number density of SM Higgs doublets (counting antiparticle states seperately)
	\begin{equation}
	Y_{H}^{\rm eq}(z_T) = \frac{ 45 \zeta(3) }{ g_{\ast} \pi^4 }.
	\end{equation}
\\
Due to their gauge interactions, the scalar triplets can annihilate into SM fermions, Higgs Bosons, and gauge bosons $T \bar{T} \to L \bar{L}, H \bar{H}, AA, YY$. The overall annihilation rate $\gamma_{\mathsmaller{TA}}(z_{T})$ can be found using the reduced cross sections reported in~\cite[Eqs.~(16a)-(16d)]{Hambye:2005tk}.  Sommerfeld enhancement is taken into account by following~\cite{Strumia:2008cf}. The gauge couplings are found at the appropriate scale, $\sim T$, using the RGEs in~\cite{Buttazzo:2013uya}. The off-shell washout rates can be found by making use of the reduced cross sections in~\cite[Eqs.~(17a) and (17b)]{Hambye:2005tk}, with appropriate modifications to make use of the RIS subtracted propagator, using the procedure of~\cite{Giudice:2003jh}, so that $\hat{\sigma}_{Ts} \to \hat{\sigma}_{Ts}^{\rm sub}$. (We also correct some trivial typos by replacing $M_T \to \tilde{m}_{T}$ inside the square brackets in~\cite[Eqs.~(17a) and (17b)]{Hambye:2005tk}. Then we identify $\tilde{m}_{T}$ with $\tilde{m}_{1}$, in our notation. We also set the parameter $m_{H}=0.05 \; \mathrm{eV}$ ---  it is the contribution to the neutrino mass mediated by heavier particles, not the SM Higgs mass. The $M_{T}^2$ factor outside the square brackets in~\cite[Eqs.~(17a) and (17b)]{Hambye:2005tk} is of course simply $M_{1}^2$ in our notation.)

 The total off-shell rate is then captured by
	\begin{equation}
	\gamma_{T}^{\rm sub}(z_T) =  \frac{1}{64\pi^{4} } \int ds s^{1/2} ( \hat{\sigma}_{T s}^{\rm sub} + \hat{\sigma}_{T t} ) K_{1}\left( \frac{ \sqrt{s} z_{T} }{ M_1 } \right).
	\end{equation}
Having collected the above definitions and rates we can now report the Boltzmann equations for type-II leptogenesis with density perturbations. Combining the results of~\cite{Hotokezaka:2025ewq,Hambye:2003rt}, these are given by:
	\begin{widetext}
	\begin{align}
	[1 - \Psi(\bar{z})] s(z_{T})H(\bar{z})\bar{z} \frac{ d \Sigma_{T} } { d \bar{z} } & = -  \gamma_{T}(z_{T}) \left[ \frac{ \Sigma_{T} }{ \Sigma_{T}^{\rm eq}(z_T) } - 1 \right]- 2 \gamma_{\mathsmaller{TA}}(z_{T}) \left[ \left( \frac{\Sigma_{T}}{\Sigma_{T}^{\rm eq}(z_T) } \right)^2 - 1 \right], \\
	[1 - \Psi(\bar{z})] s(z_{T})H(\bar{z})\bar{z} \frac{ d \Delta_{L} } { d \bar{z} } & =  \epsilon_{T} \gamma_{T}(z_{T}) \left[ \frac{ \Sigma_{T} }{ \Sigma_{T}^{\rm eq}(z_T) } - 1 \right] - 2 \gamma_{T}^{\rm sub}(z_T) \left[ \frac{ \Delta_{L} }{ Y_{L}^{\rm eq} } + \frac{ \Delta_{H} }{ Y_{H}^{\rm eq} }  \right]   - 2 B_L \gamma_{T}(z_{T}) \left[ \frac{ \Delta_{L} }{ Y_{L}^{\rm eq} } + \frac{ \Delta_{T} }{ \Sigma_{T}^{\rm eq}(z_{T}) } \right], \nonumber \\
		[1 - \Psi(\bar{z})] s(z_{T})H(\bar{z})\bar{z} \frac{ d \Delta_{H} } { d \bar{z} } & =  \epsilon_{T} \gamma_{T}(z_{T}) \left[ \frac{ \Sigma_{T} }{ \Sigma_{T}^{\rm eq}(z_T) } - 1 \right] - 2 \gamma_{T}^{\rm sub}(z_T) \left[ \frac{ \Delta_{L} }{ Y_{L}^{\rm eq} } + \frac{ \Delta_{H} }{ Y_{H}^{\rm eq} }  \right]   - 2 B_H \gamma_{T}(z_{T}) \left[ \frac{ \Delta_{H} }{ Y_{H}^{\rm eq} } - \frac{ \Delta_{T} }{ \Sigma_{T}^{\rm eq}(z_{T}) } \right]. \nonumber
	\end{align}
The main differences to the type-I case is that there is now the gauge annihilation term, $\gamma_{\mathsmaller{TA}}(z_{T})$, acting to keep $\Sigma_{T}$ close to its equilibrium value, and there are two asymmetries to track. The final baryon asymmetry is given by $  Y_{\mathcal{B}} = (28/79) \, Y_{\mathcal{B}-\mathcal{L}} = -(28/79) \, \Delta_{L}$. \\
	\end{widetext}

\begin{figure*}[t]
\begin{center}
\includegraphics[width=0.32 \textwidth]{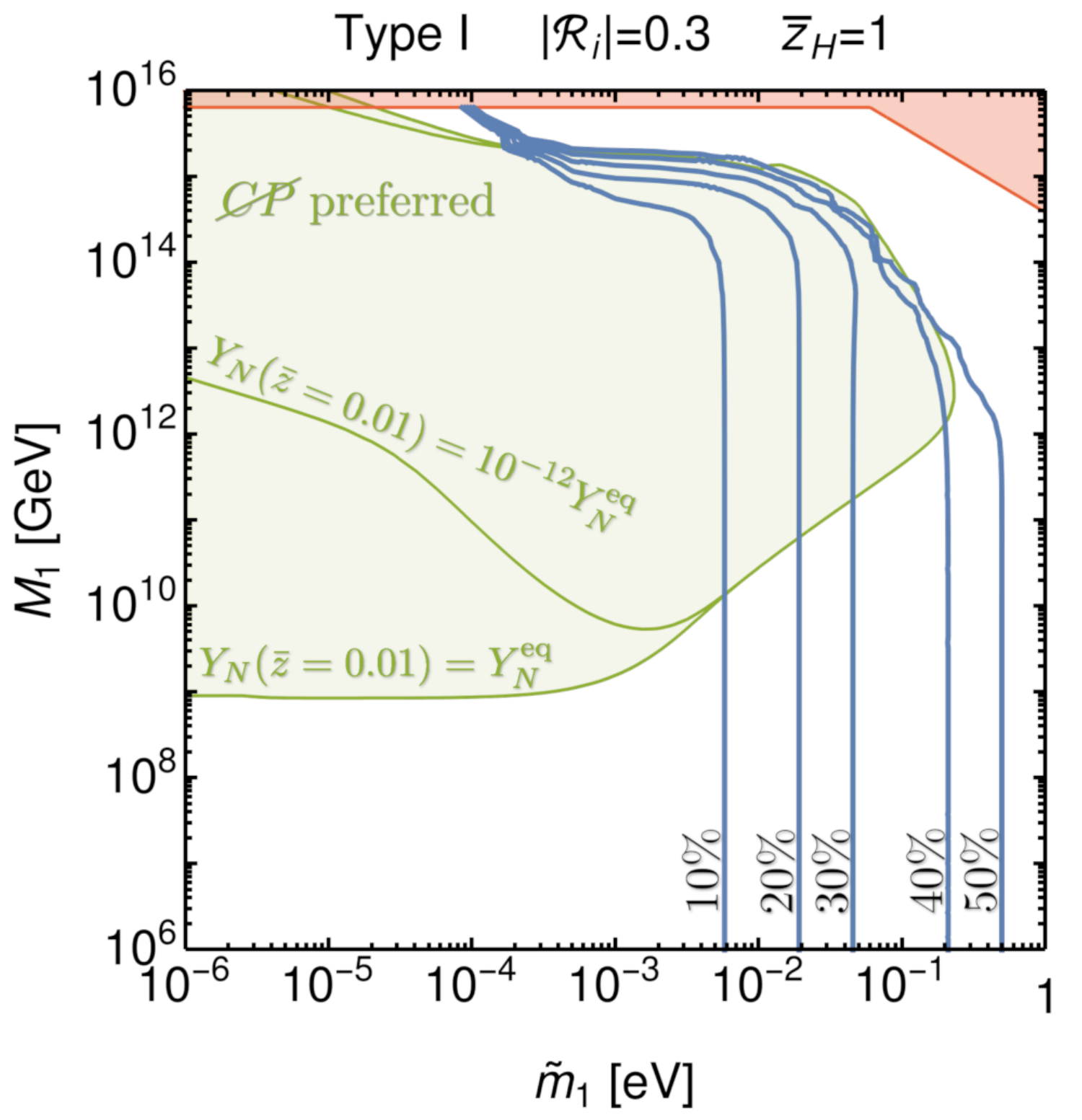} \,
\includegraphics[width=0.32 \textwidth]{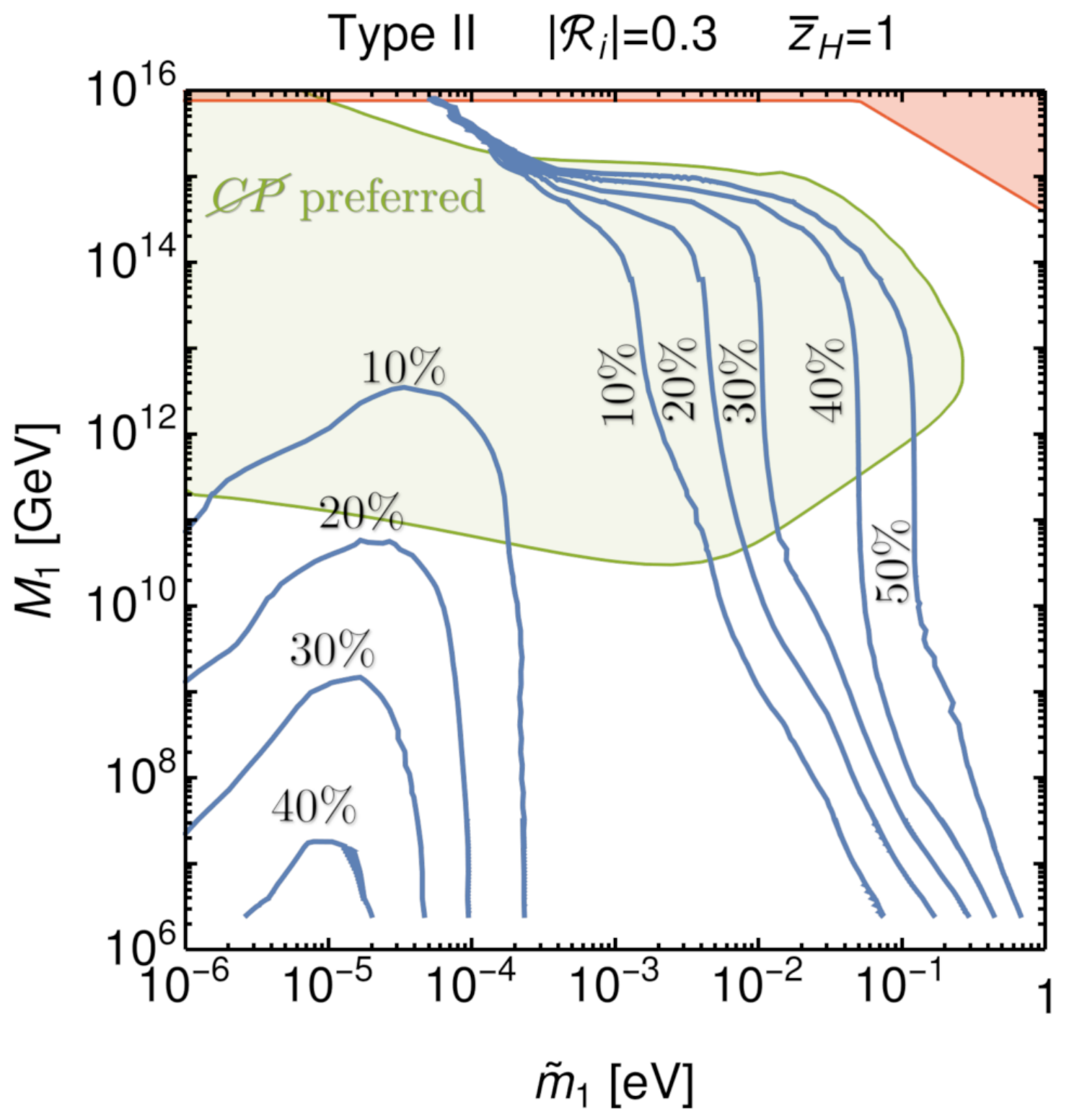} \,
\includegraphics[width=0.32 \textwidth]{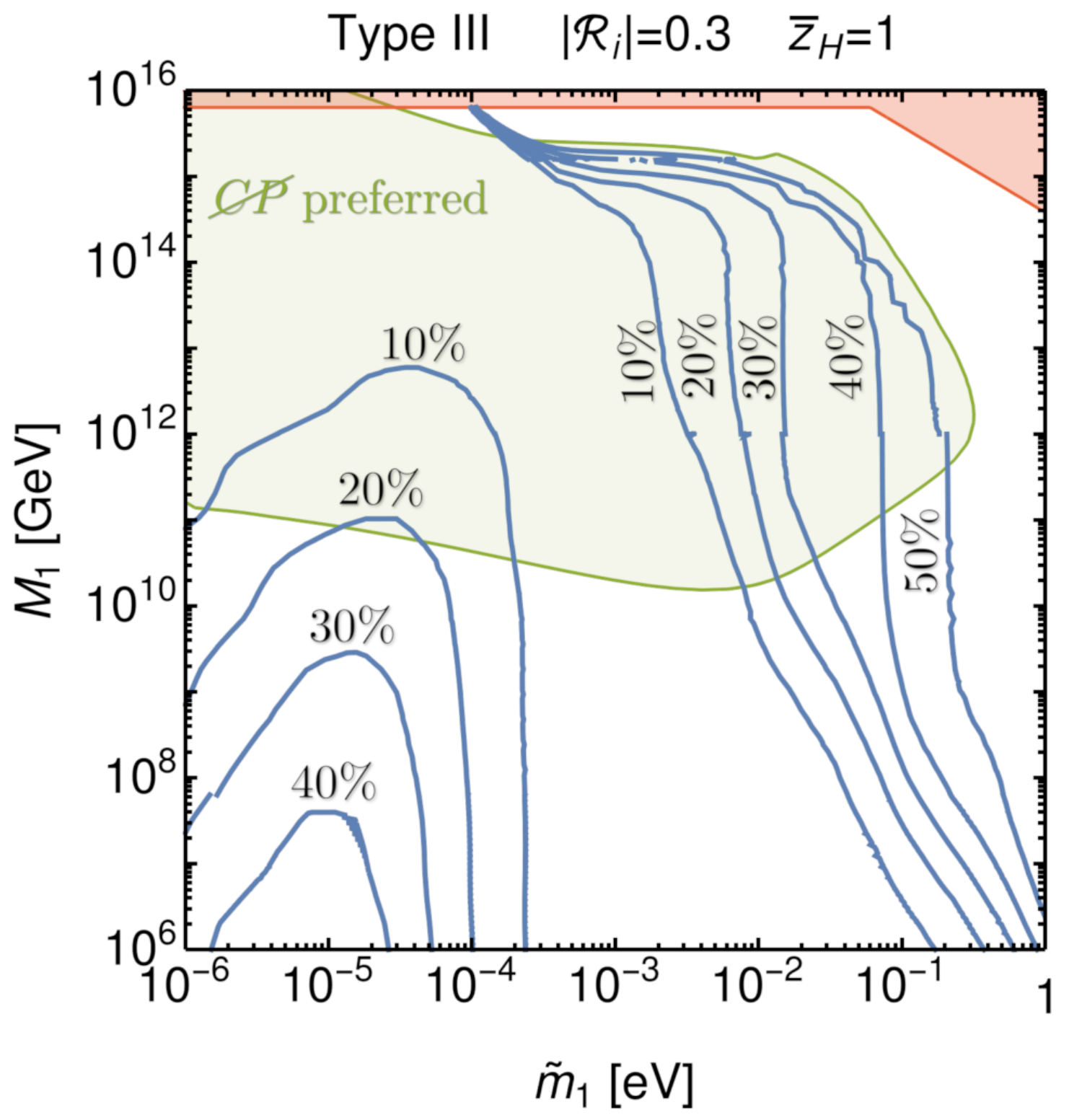}
\end{center}
\caption{
Contours showing the enhancement of the yield with perturbations for the three seesaw models (with $B_L=B_H=0.5$ for type-II).  The new result shows that in the presence of perturbations, i.e. acoustically driven freeze-out, gauge interactions allow yield enhancements in the weak washout regime, $\tilde{m_{1}} < m_{\ast}$. This occurs in parameter ranges where the decay time approximately matches the freeze-out time of the gauge interactions. As well established in the previous literature, preferred regions (green) assuming hierarchical spectrum CP violation limits, Eqs.~\eqref{eq:cptype1}-\eqref{eq:cptype3}, are at somewhat higher $M_{1}$ in the type-II and type-III scenarios, due to the gauge interactions keeping the number densities closer to equilibrium. The green regions are shown without the perturbations. The increase in the yield, due to the perturbations, will only lead to small changes at the borders of these areas as can be inferred by the blue contours. (The assumption of the initial $N$ abundance for the type-I seesaw affects the preferred region at small $\tilde{m}_{1}$.) At large values of $M_{1}$ the yields become suppressed due to off-shell $2 \to 2$ washout interactions. For very large $M_{1}$ the couplings become non-perturbative (red shaded regions).  }
\label{fig:enhancements}
\end{figure*}		

\subsection{Type-III Boltzmann equation} 

For type-III leptogenesis we are interesting in tracking the fermion triplet number density
	\begin{equation}
	Y_{\Sigma} = \frac{ n_{\Sigma} }{ s },
	\end{equation}
with equilibrium number density 
	\begin{equation}
	Y_{\Sigma}^{\rm eq}(z_T) = \frac{ n_{\Sigma }^{\rm eq} }{ s } = \frac{ g_{\Sigma} M_{1}^3  }{ 2 \pi^2 z_T s(z_T)} K_{2}(z_{T}),
	\end{equation}
where $g_{\Sigma } = 6 $. The thermally averaged decay rate times number density is given by
	\begin{equation}
	\gamma_{\Sigma}(z_{T}) = \frac{ K_{1}(z_{T}) }{ K_{2}(z_{T}) } n_{\Sigma }^{\rm eq}(z_{T}) \Gamma_{\Sigma} . 
	\end{equation}
The number density is suppressed by annihilations into gauge bosons and SM fermion doublets, $\Sigma \bar{\Sigma} \to AA, L\bar{L}$ with a rate denoted $\gamma_{\Sigma A}(z_{T})$ and a reduced cross section $\hat{\sigma}_{\Sigma A}$ reported in~\cite[Eq.~(29c)]{Hambye:2003rt}. Sommerfeld enhancement is again taken into account following~\cite{Strumia:2008cf}.

We also require the off-shell washout rates for processes mediated by a $\Sigma$ propagator,  $LH \to \bar{H}\bar{L}$ and $LL \to \bar{H}\bar{H}$, using
		\begin{equation}
	\gamma_{\Sigma}^{\rm sub}(z_T) =  \frac{1}{64\pi^{4} } \int ds s^{1/2} ( \hat{\sigma}_{\Sigma s}^{\rm sub} + \hat{\sigma}_{\Sigma t} ) K_{1}\left( \frac{ \sqrt{s} z_{T} }{ M_{1} } \right),
	\end{equation}
where the two cross sections can be found in~\cite[Eqs.~(29a) and (29b)]{Hambye:2003rt}. (We also corrected two sign errors therein, which can be found by comparison to the analogous cross sections in~\cite[Eqs.~(92) and (93)]{Giudice:2003jh}, and one of which otherwise gives unphysical increases in $\gamma_{\Sigma}^{\rm sub}(z_T)$ at low temperatures.)

Having collected the required rates, we can now give the Boltzmann equations with the inclusion of the density perturbations for type-III leptogenesis. Blending the results of~\cite{Hotokezaka:2025ewq,Hambye:2005tk,Strumia:2008cf}, these are:
\begin{widetext}
\begin{align}
[1 - \Psi(\bar{z})] s(z_{T})H(\bar{z})\bar{z} \frac{ d Y_{\Sigma} } { d \bar{z} } & = -  \gamma_{\Sigma}(z_{T}) \left[ \frac{Y_{\Sigma}}{Y_{\Sigma}^{\rm eq}(z_T) } - 1 \right]- 2 \gamma_{\Sigma A}(z_{T}) \left[ \left( \frac{Y_{\Sigma}}{Y_{\Sigma}^{\rm eq}(z_T) } \right)^2 - 1 \right], \\
[1 - \Psi(\bar{z})] s(z_{T})H(\bar{z})\bar{z} \frac{ d Y_{\mathcal{B}-\mathcal{L}} } { d \bar{z} } & = - \epsilon_{\Sigma} \gamma_{\Sigma}(z_{T}) \left[ \frac{Y_{\Sigma}}{Y_{\Sigma}^{\rm eq}(z_T) } - 1 \right] - \frac{ Y_{\mathcal{B}-\mathcal{L}} }{ Y_{L}^{\rm eq}(z_T) } \left[ \frac{ \gamma_{\Sigma}(z_T) }{2} + 2 \gamma_{\Sigma}^{\rm sub}(z_T)   \right].
\end{align}
As can be readily seen, up to to the annihilation term and some factors of three arising from the multiplicity of the $\Sigma$, hidden in the number densities and cross sections, the type-III Boltzmann equations are very similar to the type-I scenario. \\
\end{widetext}  

\section{Results}

Having the required Boltzmann equations we then solve them numerically in order to find the yield of baryons with and without the density perturbations. Example solutions for all the leptogenesis types are shown in Fig.~\ref{fig:bmannexamples}. As can be seen, there is a general tendency of the freezing out particle to stay overabundant above the average of the oscillation, which typically leads to an enhancement of the generated asymmetry.

This is then confirmed by averaging over the choice of phase $\delta$ appearing through Eqs.~\eqref{eq:Psifunc} and \eqref{eq:deltafunc}, which corresponds also to a spatial average, as different points in the universe will experience different phases of the oscillation (as explained in~\cite{Hotokezaka:2025ewq}). After taking the $\delta$ average, in Fig.~\ref{fig:weakvsstrong} we show the dependence of the yield as a function of $|\mathcal{R}_{i}|$ for the three seesaws and contrast the qualitative differences between the seesaws in the weak and strong washout regimes (this also confirms the results for type-I of~\cite{Hotokezaka:2025ewq}).

Finally in Fig.~\ref{fig:enhancements}, we collect our results, by showing standard plots of the preferred regions for leptogenesis in the different seesaw models (without the perturbations), together with contours showing the enhancement of the yield for acoustically driven freeze-out, in terms of the parameters $M_{1}$ and $\tilde{m_{1}}$. Cosmological observations disfavour $\sum m_{\nu i} > 0.12 \; \mathrm{eV}$~\cite{Planck:2018vyg,eBOSS:2020yzd}, so also $\tilde{m}_{1} > 0.12 \; \mathrm{eV}$ is disfavoured~\cite{Buchmuller:2003gz}, but we extend the plots to higher values for completeness. At large values of $M_{1}$ the couplings required to explain the neutrino masses become non-perturbative. Observations of the CMB limit the post-inflationary reheat temperature to $T_{\rm RH} < 6.6 \times 10^{15} \; \mathrm{GeV}$~\cite{Planck:2018jri,BICEP:2021xfz} which gives a similar consistency constraint on the scenario, $M_{1} < T_{\rm RH}$, as perturbativity. 

As initial conditions we typically assume thermal abundance of the heavy see-saw states and zero initial asymmetry. For the type-I seesaw, the final asymmetry is sensitive to both these initial conditions in the weak washout regime. We show in Fig.~\ref{fig:enhancements} the effect of assuming a highly suppressed abundance, $Y_{N}(\bar{z}=0.01) = 10^{-12} \, Y_{N}^{\rm eq} $, on the preferred region for hierarchical spectra (relevant at small $\tilde{m_{1}}$). The sensitivity is outside the parameter region in which the acoustically driven freeze-out plays a role. For type-II and type-III, as the gauge interactions bring the heavy states to their thermal abundance values, there is no dependence on the assumed initial abundance of the heavy state (only depedence on the assumed zero initial asymmetry in the weak washout regime).

Note we take $B_H=B_L=0.5$ for type-II only for simplicity. Away from this limit, the washout rate decreases, but so does the amount of CP violation, so that the baryon yield is suppressed. (There are also some differences in the preferred green region from CP violation at large $M_{1}$ and $\tilde{m}_{1}$ for type-II, compared with the results of~\cite[Fig. 2]{Strumia:2008cf}, we were not able to find the source of this discrepancy, e.g.~by allowing for larger $\sum m_{\nu i}^{2}$ in Eq.~\eqref{eq:cptype2}, the bound on $|\epsilon_{T}|$.)

In the strong washout regime, $\tilde{m_{1}} > m_{\ast}$, departure from equilibrium is set by the decay rate, making the three scenarios qualitatively similar. In the weak  washout regime, $\tilde{m_{1}} < m_{\ast}$, freeze-out of the gauge annihilations plays the central role in determining the departure from equilibrium in type-II and type-III, leading to significant qualitative differences between these scenarios and type-I leptogenesis. This was already known to be the case for the behaviour of the evolution of the particle number densities and asymmetries without the density perturbations. The effects of the acoustically driven freeze-out follow a similar pattern.

\section{Conclusion}

We have extended the study of acoustically driven freeze out to include $2 \to 2$ interactions, by examining in some detail leptogenesis in the type-II and type-III seesaw models in the presence of perturbations, and contrasting our results with those of type-I leptogenesis. Modest enhancements of the yield, $\sim 40 \%$, are possible with primordial curvature perturbations $\mathcal{R}_{i} \sim 0.3$, already in the weak washout regime in the scenarios with gauge interactions. (The crucial ingredient being annihilations involving the species freezing out in the external states --- similar effects could be expected in extensions of the type-I seesaw involving annihilations of $N_{1}$ into scalars~\cite{AristizabalSierra:2014uzi}.)

Assuming the existence of such primordial small scale perturbations, the size of the effect is of the same order as the effect of the Sommerfeld enhancment in type-II and III leptogenesis~\cite{Strumia:2008cf} (which we have also included in our calculation), although most relevant for somewhat differing areas of parameter space. 

Other sizable corrections we have not included are the following: If the heavy states have small mass splittings, resonance effects can lead to significant increases in the CP violation, and hence lower the allowed regions to the $\sim$ TeV scale in both the type-I~\cite{Pilaftsis:2003gt} and type-II/III cases~\cite{Strumia:2008cf}. Flavour effects can also be significant, improving the efficiency by around an order of magnitude, and thus lowering the minimum seesaw scale for hierarchical leptogenesis in all seesaw cases also by around an order of magnitude~\cite{Abada:2006fw,Nardi:2006fx,Abada:2006ea,Lavignac:2015gpa}. Spectator processes, not included here, which are in thermal equilibrium during leptogenesis lead to $\mathcal{O}(1)$ differences in the yield~\cite{Buchmuller:2001sr,Nardi:2005hs}.

For the range of effective neutrino masses considered here, indeed for $\tilde{m}_{1} \gtrsim 10^{-7.5}$ eV, the heavy neutrinos do not come to dominate the universe before their decay (even in the type-I case assuming an initial thermal abundance), so there is no early epoch of matter domination. Once well inside the Horizon and well after asymmetry generation~\cite{Jeong:2014gna,Nakama:2014vla}, however, diffusion damping of the oscillations will lead to entropy production, at second order in the perturbations~\cite{Chluba:2012gq,Chluba:2012we}. This increases the entropy density and thus leads to a suppression of the yield by a factor $s_{f}/s_{i} = (1 + 3 \langle \delta_{T} ^{2} \rangle $). For our monochromatic perturbations, the time and $\delta$ phase average is  $\langle \delta_{T} ^{2}\rangle = |\mathcal{R}_{i}|^{2}/4$, so the correction is $< 7\%$ for the perturbations $|\mathcal{R}_{i}| \leq 0.3 $ considered here. (Larger effects are possible for broad, scale invariant enhancements of the perturbation spectra at small scales~\cite{Jeong:2014gna,Nakama:2014vla,Inomata:2016uip}.)

It would be of interest to study acoustically driven freeze-out in the context of dark matter, to see what effect, if any, is present in the freeze-out of stable species. Preliminary investigations suggest that in the standard DM freeze-out scenario, modest enhancements of the required DM annihilation cross section $\sim 10 \%$ are necessary to match the relic abundance, for perturbations $\mathcal{R}_{i} \sim 0.3$~\cite{baldes2} (for effects of small scale perturbations during freeze-in, see \cite[Fig.~1]{Stebbins:2023wak}). It would also be of interest to see whether models featuring inflation, or strong early universe phase transitions, can be found which link physics of the particles undergoing freeze-out, with the generation of large $\mathcal{R}_{i}$ on the length scales $\sim 1/H(\bar{T}=M)$, relevant for acoustically driven freeze-out.      

\subsection*{Acknowledgements}
This work was supported by the European Union's Horizon 2020 research and innovation programme under Grant Agreement No. 101002846, ERC CoG ``CosmoChart."

\bibliography{LGEN}
\end{document}